\documentclass[12pt]{article}

\usepackage{latexsym}

\usepackage{graphicx}

\begin{document}


\title{5D Einstein-Maxwell solitons and  concentric rotating dipole black rings}

\author{
     Stoytcho S. Yazadjiev \thanks{E-mail: yazad@phys.uni-sofia.bg}\\
{\footnotesize  Institut f\"ur Theoretische Physik, Universit\"at G\"ottingen,}\\
{\footnotesize  Friedrich-Hund-Platz 1, D-37077 G\"ottingen, Germany}\\
{\footnotesize and} \\
{\footnotesize  Department of Theoretical Physics,
                Faculty of Physics, Sofia University,}\\
{\footnotesize  5 James Bourchier Boulevard, Sofia~1164, Bulgaria }\\
}

\date{}

\maketitle

\begin{abstract}
We discuss the application of the solitonic techniques to the 5D Einstein-Maxwell gravity.
As an illustration we construct an exact solution describing two concentric rotating dipole black
rings. The properties of the solution are investigated.
\end{abstract}


\sloppy

\section{Introduction}

The black holes in higher dimensions  are attracting a lot of interest and now the higher dimensional gravity
is a rapidly growing area of research. The black hole solutions in higher dimensions
demonstrate very interesting properties and features absent in four dimensions. In spacetimes with number of dimensions greater than four the black objects are no longer uniquely  determined  by the conserved asymptotic charges. A lot of research was devoted to the black holes in five dimensions where   many interesting and important from theoretical point of view exact solutions were found and analyzed \cite{ER2}-\cite{EvslinKrishnan1}. The accumulation of exact solutions naturally raises the question of the classification of the black solutions. Some basic results in the classification of the black holes solutions in five dimensions were recently obtained in \cite{HolandsYazadjiev1} and \cite{HolandsYazadjiev2} for vacuum Einstein and  a certain sector of the Einstein-Maxwell gravity. It is worth noting that as a byproduct of the horizon topology classification theorem given in \cite{HolandsYazadjiev1} a new type of possible asymptotically flat 5D black objects was predicted--the so-called black lenses i.e. black objects whose horizon topology is that of a lens space. The black lenses must have two angular momenta and most probably could be constructed
by using solitonic techniques.

In the present paper we discuss the application of the solitonic techniques in the 5D Einstein-Maxwell gravity.
For the vacuum case the application of the solitonic techniques to the 5D stationary, axi-symmetric gravity was discussed in \cite{Pomeransky}--\cite{IgichiMishima1}.
It was shown in \cite{Yazadjiev3} that there is a completely integrable sector in 5D Einstein-Maxwell gravity and the corresponding Lax pair was explicitly found which opens the way to applying the inverse-scattering method \cite{BZ1}. This sector namely will be considered here in the context of the solitonic Backlund transformations. Following \cite{Yazadjiev3} we reduce the stationary, axi-symmetric Einstein-Maxwell equations to two effective four-dimensional problems. More precisely the 5D Einstein-Maxwell equations are reduced to two Ernst equations which allows us  to apply the solitonic techniques to the considered sector of the 5D Einstein-Maxwell gravity. As an explicit illustration of the solitonic techniques we construct an exact solution to the 5D Einstein-Maxwell equations describing concentric rotating dipole black rings.

\section{Solitonic transformations }

In five dimensions the Einstein-Maxwell  equations read
\begin{eqnarray}\label{EMFE}
&&R_{\mu\nu} = {1\over 2} \left(F_{\mu\lambda}F_{\nu}^{\,\lambda}
 - {1\over 6} F_{\sigma\lambda}F^{\sigma\lambda} g_{\mu\nu}\right),  \\
&&\nabla_{\mu} F^{\mu\nu} = \nabla_{[\mu}F_{\nu\lambda ]} =0  \nonumber.
\end{eqnarray}

In this paper, following \cite{Yazadjiev3}  we consider 5D EM  gravity in spacetimes with  three commuting Killing vectors:
one timelike Killing vector $T={\partial \over \partial t}$ and two spacelike Killing vectors $K_{\psi}={\partial \over \partial \psi}$ and $K_{\phi}={\partial \over \partial \phi}$.  We also assume
that the Killing vector $K_{\phi}={\partial \over \partial \phi}$ is hypersurface orthogonal.
In this case the 5D spacetime metric can be written in the form
\begin{eqnarray}
ds^2= - e^{2\chi-u}\left(dt - \omega d\psi \right)^2 + e^{-2\chi-u}\rho^2 d\psi^2 + e^{-2\chi-u} e^{2\Gamma}\left(d\rho^2 + dz^2 \right) +
e^{2u}d\phi^2
\end{eqnarray}
where all the metric functions depend on the canonical coordinates $\rho$ and $z$ only.

It is useful to introduce the so-called twist potential $f$ defined by the equations
\begin{eqnarray}
&&\partial_{\rho}f= -{1\over 2}{e^{4\chi}\over \rho}\partial_{z}\omega ,\\
&&\partial_{z}f= {1\over 2}{e^{4\chi}\over \rho}\partial_{\rho}\omega .
\end{eqnarray}
We also make the following (local) ansatz  for the gauge potential 1-form $A$, namely

\begin{equation}
A=A_{\phi}d\phi.
\end{equation}

Further we introduce the two complex Ernst potentials ${\cal E}_{1}$ and ${\cal E}_{2}$ defined by
\begin{eqnarray}
&&{\cal E}_{1}= e^{u} + {i\over \sqrt{3}} A_{\phi},\\
&&{\cal E}_{2}= e^{2\chi} + 2if .
\end{eqnarray}

With the help of the Ernst potentials the dimensionally reduced 5D Einstein-Maxwell equations can be written in the following form
\begin{eqnarray}
&&\left({\cal E}_{1} + {\cal E}^{*}_{1}\right)\left(\partial^2_{\rho} {\cal E}_{1} + \rho^{-1} \partial_{\rho}{\cal E}_{1}
+ \partial^2_{z} {\cal E}_{1}\right)= 2 \left(\partial_{\rho}{\cal E}_{1}\partial_{\rho}{\cal E}_{1}
+ \partial_{z}{\cal E}_{1}\partial_{z}{\cal E}_{1} \right), \nonumber
\end{eqnarray}
\begin{eqnarray}
&&\left({\cal E}_{2} + {\cal E}^{*}_{2}\right)\left(\partial^2_{\rho} {\cal E}_{2} + \rho^{-1} \partial_{\rho}{\cal E}_{2}
+ \partial^2_{z} {\cal E}_{2}\right)= 2 \left(\partial_{\rho}{\cal E}_{2}\partial_{\rho}{\cal E}_{2}
+ \partial_{z}{\cal E}_{2}\partial_{z}{\cal E}_{2} \right),
\end{eqnarray}
\begin{eqnarray}
&&\rho^{-1}\partial_{\rho}\Gamma= {1\over \left({\cal E}_{2} + {\cal E}^{*}_{2} \right)^2}
\left(\partial_{\rho}{\cal E}_{2}\partial_{\rho}{\cal E}^{*}_{2} - \partial_{z}{\cal E}_{2}\partial_{z}{\cal E}^{*}_{2} \right)
+ {3\over \left({\cal E}_{1} + {\cal E}^{*}_{1} \right)^2}
\left(\partial_{\rho}{\cal E}_{1}\partial_{\rho}{\cal E}^{*}_{1} - \partial_{z}{\cal E}_{1}\partial_{z}{\cal E}^{*}_{1} \right),\nonumber \\
\nonumber \\
 &&\rho^{-1}\partial_{z}\Gamma= {2\over \left({\cal E}_{2} + {\cal E}^{*}_{2} \right)^2} \partial_{\rho}{\cal E}_{2} \partial_{z}{\cal E}^{*}_{2}
+ {6\over \left({\cal E}_{1} + {\cal E}^{*}_{1} \right)^2} \partial_{\rho}{\cal E}_{1} \partial_{z}{\cal E}^{*}_{1} .\nonumber
\end{eqnarray}

In this way we reduced the problem of solving the 5D EM equations to two effective 4D problems i.e. two Ernst equations.  The central and most difficult task is to solve the nonlinear Ernst equations. Fortunately there are mathematical techniques developed in the 4D gravity which can be applied to the Ernst equation. Here we apply the solitonic  Neugebauer \cite{Neugebauer1} or Hoenselaers-Kinnersley-Xantopoulos \cite{HKX1} transformation (see also \cite{SKMCHH}).

In order to facilitate the reader we shall  briefly describe the Neugebauer 2N-soliton transformation.
More details can be found in \cite{Neugebauer1},\cite{HKX1},\cite{SKMCHH}. Let us consider a real seed solution ${\cal E}^{0}$ of the Ernst equation given by
\begin{eqnarray}
{\cal E}^{0}= e^{2\chi_0}
\end{eqnarray}
where $\chi_{0}$ is a harmonic function of the canonical coordinates $\rho$ and $z$. Then a new solution to the Ernst equation corresponding to adding 2N solitons to the seed background  is given by
\begin{eqnarray}
{\cal E}= e^{2\chi_{0}} {\det\left({\alpha_{p}R_{k_p} - \alpha_{q}R_{k_q} \over k_p - k_q } -1 \right)\over
\det\left({\alpha_{p}R_{k_p} - \alpha_{q}R_{k_q} \over k_p - k_q } +1 \right)}
\end{eqnarray}
where
\begin{eqnarray}
R_{k_p}= \sqrt{\rho^2 + (z-k_p)^2},
\end{eqnarray}
$k_p$ are real constants and $p=1,3,...,2N-1$, $q=2,4,...,2N$. The functions $\alpha_{p}$ are defined by
\begin{eqnarray}
\alpha_{p}= {\mu_p + ie^{2\Phi_p}\over \mu_p - ie^{2\Phi_p}}
\end{eqnarray}
where $\mu_p$ are constants and $\Phi_p$ is a solution of the Riccati equation
\begin{eqnarray}
d\Phi_p= {1\over 2} \sqrt{k_p - i\zeta^{*} \over k_p + i\zeta}  \partial_{\zeta} \ln{\cal E}^{0} d\zeta  +
{1\over 2}\sqrt{k_p +i\zeta  \over k_p - i\zeta^{*}}  \partial_{\zeta^{*}}\ln{\cal E}^{0}d\zeta^{*}
\end{eqnarray}
with $\zeta= \rho + i z$.

In this paper we will consider seed potentials of the form
\begin{eqnarray}\label{seed}
\chi_{0}= \sum_{i} \varepsilon_i {\tilde U}_{\nu_i}
\end{eqnarray}
where $\varepsilon_i$ and $\nu_i$ are constants and ${\tilde U}_{\nu_i}$ is defined by
\begin{eqnarray}
{\tilde U}_{\nu_i}={1\over 2} \ln\left[\sqrt{\rho^2 +(z-\nu_i)^2} + (z-\nu_i)\right]={1\over 2} \left[R_{\nu_i}+ (z-\nu_i)\right].
\end{eqnarray}

For the described seed function the solution of the Riccati equation is
\begin{eqnarray}
\Phi_{p} = \sum_{i} {\varepsilon_i\over 2} \ln\left({e^{2U_{k_p}} + e^{2{\tilde U}_{\nu_i}}\over e^{{\tilde U}_{\nu_i}} }\right)
\end{eqnarray}
where $U_{k_p}$ is defined by
\begin{eqnarray}
U_{k_p}= {1\over 2}\ln\left[R_{k_p} -(z-k_p)\right] .
\end{eqnarray}

Here we will be interested in the 2-soliton transformation.  The 2-soliton Neugebauer transformation applied to the seed ${\cal E}^{0}=e^{2\chi_{0}}$ is equivalent to the  2 rank-zero Hoenselaers-Kinnersley-Xantopoulos transformations  applied however to the seed ${\tilde {\cal E}}^{0}=e^{2{\tilde \chi}_{0}}$ where
\begin{eqnarray}
{\tilde \chi}_{0} = \chi_{0} +{1 \over 2} \ln{R_{k_1}+ R_{k_2} - \Delta k\over R_{k_1} + R_{k_2} +\Delta k}
\end{eqnarray}
and $\Delta k=k_1 - k_2$.

Once having  solutions to the Ernst equations we must find the metric function $\Gamma$. In order to do so it is convenient to introduce the auxiliary functions ${\hat \Gamma_{i}}$ defined by the equations
\begin{eqnarray}
&&\rho^{-1}\partial_{\rho}{\hat \Gamma_{i}}= {1\over \left({\cal E}_{i} + {\cal E}^{*}_{i} \right)^2}
\left(\partial_{\rho}{\cal E}_{i}\partial_{\rho}{\cal E}^{*}_{i} - \partial_{z}{\cal E}_{i}\partial_{z}{\cal E}^{*}_{i} \right), \nonumber \\
\\
 &&\rho^{-1}\partial_{z}{\hat \Gamma_{i}}= {2\over \left({\cal E}_{i} + {\cal E}^{*}_{i} \right)^2} \partial_{\rho}{\cal E}_{i} \partial_{z}{\cal E}^{*}_{i} .\nonumber
\end{eqnarray}
As a consequence we have
\begin{eqnarray}
\Gamma= {\hat \Gamma_{2}} + 3{\hat \Gamma_{1}}.
\end{eqnarray}

The functions ${\hat \Gamma}$ can be found as follows. Consider a solution ${\cal E}$ to the Ernst equation corresponding to the seed $e^{2\chi_{0}}$ such that
\begin{eqnarray}
Re {\cal E}= e^{2\chi_{0}} {A\over B }.
\end{eqnarray}
Then the function ${\hat \Gamma}$ is given by
\begin{eqnarray}
e^{2{\hat \Gamma}}= C {A e^{2H}\over \left(R_{k_{1}} +  R_{k_{2}}\right)^2 - (\Delta k)^2 },
\end{eqnarray}
where $C$ is a constant and $H$ is a solution to the linear system
\begin{eqnarray}
&&\rho^{-1}\partial_{\rho}H= \left(\partial_{\rho}{\tilde \chi_{0}}\right)^2
- \left(\partial_{z}{\tilde \chi_{0}}\right)^2, \nonumber \\ \\
&&\rho^{-1}\partial_{z}H=2 \partial_{\rho}{\tilde \chi_{0}} \partial_{z}{\tilde \chi_{0}}.\nonumber
\end{eqnarray}

For seed functions of the form (\ref{seed}) the function $H$ is given by
\begin{eqnarray}
H = \gamma_{k_1,k_1} -2\gamma_{k_1,k_2} + \gamma_{k_2,k_2}  + 2\sum_{i}\left(\gamma_{k_1,\nu_i}-\gamma_{k_2,\nu_i} \right) +
\sum_{i,j}\varepsilon_i \varepsilon_j \gamma_{\nu_i,\nu_j}
\end{eqnarray}
where
\begin{eqnarray}
\gamma_{k,l}= {1\over 2}{\tilde U}_{k} + {1\over 2}{\tilde U}_{l} -{1\over 4} \ln\left[R_k R_l + (z-k)(z-l) +\rho^2  \right] .
\end{eqnarray}

\section{Concentric rotating dipole black rings}

In order to illustrate the application of the solitonic techniques to the 5D Einstein-Maxwell gravity we shall
construct an exact solution describing two concentric rotating dipole black rings. In order to generate such a solution we need an appropriate seed solution. The natural choice is the seed solution for the vacuum black di-ring \cite{IgichiMishima2}. This solution is given by the following metric functions
\begin{eqnarray}\label{seed}
2\chi_{0}= {\tilde U}_{\nu_1} - {\tilde U}_{\nu_{2}} + {\tilde U}_{\nu_{3}}+ {\tilde U}_{\nu_{4}}
- {\tilde U}_{\nu_{5}}, \\ \nonumber \\
u_{0}= {\tilde U}_{\nu_1} + {\tilde U}_{\nu_{2}} - {\tilde U}_{\nu_{3}}+ {\tilde U}_{\nu_{4}}
- {\tilde U}_{\nu_{5}},\nonumber
\end{eqnarray}
where $\nu_i$ are constants. At this stage the parameters $\nu_i$ are arbitrary. Physical restrictions and bounds on these parameters will be imposed a little later.

Let us first consider the Ernst equation for ${\cal E}_{2}$. This equation can be solved by the Neugebauer $2$-soliton transformation applied to the seed Ernst potential ${\cal E}^{0}_2=e^{2\chi_{0}}$.  The parameters of the solitonic transformation will be denoted by $l_1$, $l_2$, ${\hat\alpha}$ and ${\hat\beta}$.  In explicit form the solution is
given by
\begin{eqnarray}
&&e^{2\chi}= e^{2\chi_{0}} {A \over B} , \\\nonumber
\\
&&\omega= e^{-2\chi_{0}} { {\hat \omega }\over A }    + C_{\omega},
\end{eqnarray}
where  $C_{\omega}$ is a constant,
\begin{eqnarray}
A= &&\left[(R_{l_1}+R_{l_2})^2-(\Delta l)^2\right] (1+ a b)^2  + \left[(R_{l_1}-R_{l_2})^2-(\Delta l)^2\right](a-b)^2, \nonumber \\ \nonumber \\
B= &&\left[(R_{l_1}+R_{l_2}+\Delta l)+(R_{l_1}+R_{l_2}-\Delta l)a b \right]^2  \\
 &&+\left[(R_{l_1}-R_{l_2}-\Delta l)a - (R_{l_1}-R_{l_2}+\Delta l)b \right]^2 , \nonumber \\ \nonumber  \\
{\hat \omega} =&& [(R_{l_1} + R_{l_2})^2-(\Delta l)^2](1+a b)\left[(R_{l_1}-R_{l_2} + \Delta l)b +
(R_{l_1}-R_{l_2} - \Delta l)a\right] \nonumber \\&& \,+ [(R_{l_1} - R_{l_2})^2-(\Delta l)^2](b-a)
\left[(R_{l_1} + R_{l_2} + \Delta k) - (R_{l_1} + R_{l_2} -\Delta l)ab\right], \nonumber
\end{eqnarray}
and the functions $a$ and $b$ are defined by
\begin{eqnarray}
a = {\hat\alpha} \, {e^{2U_{l_1}} + e^{2{\tilde U}_{\nu_1}} \over e^{{\tilde U}_{\nu_1}} }
{e^{2U_{l_1}} + e^{2{\tilde U}_{\nu_2}} \over e^{{\tilde U}_{\nu_2}} }
{e^{{\tilde U}_{\nu_3}} \over e^{2U_{l_1}} + e^{2{\tilde U}_{\nu_3}}}
{e^{2U_{l_1}} + e^{2{\tilde U}_{\nu_4}} \over e^{{\tilde U}_{\nu_4}}}
{e^{{\tilde U}_{\nu_5}} \over e^{2U_{l_1}} + e^{2{\tilde U}_{\nu_5}}} ,
\end{eqnarray}
\begin{eqnarray}
b = {\hat \beta} \, {e^{{\tilde U}_{\nu_1}} \over e^{2U_{l_2}} + e^{2{\tilde U}_{\nu_1}} }
{e^{{\tilde U}_{\nu_2}} \over e^{2U_{l_2}} + e^{2{\tilde U}_{\nu_2}}}
{e^{2U_{l_2}} + e^{2{\tilde U}_{\nu_3}} \over e^{{\tilde U}_{\nu_3}} }
{e^{{\tilde U}_{\nu_4}} \over e^{2U_{l_2}} + e^{2{\tilde U}_{\nu_4}} }
{e^{2U_{l_2}} + e^{2{\tilde U}_{\nu_5}} \over e^{{\tilde U}_{\nu_5}} } .
\end{eqnarray}

In order to solve the first Ernst equation we applied the Neugebauer 2-soliton transformation to the background ${\cal E}_{1}^{0}=e^{u_{0}}$. Denoting the soliton transformation parameters by $k_1$, $k_2$, ${\hat \alpha}^\prime$ and ${
\hat\beta}^\prime$ the solution reads
\begin{eqnarray}
&&e^{u} = Re {\cal E}_{1}= e^{u_{0}} W = e^{u_{0}} {W_{1}\over W_{2}} ,\\ \nonumber \\
&&A_{\phi}= \sqrt{3}Im{\cal E}_{1}= 4\sqrt{3}\Delta k e^{u_{0}}
{\left[R_{k_1} a^\prime (1+ {b^\prime}^2) + R_{k_2} b^\prime (1+ {a^\prime}^2) \right]  \over W_2} + A^{0}_{\phi},
\end{eqnarray}
where $A^{0}_{\phi}$ is a constant and  $W_1$ and $W_2$ are given by
\begin{eqnarray}
W_1&=&\left[(R_{k_1}+R_{k_2})^2-(\Delta k)^2\right] (1+ a^\prime b^\prime)^2  + \left[(R_{k_1}-R_{k_2})^2-(\Delta k)^2\right](a^\prime-b^\prime)^2  ,\\ \nonumber \\
W_2&=&\left[(R_{k_1}+R_{k_2}+\Delta k)+(R_{k_1}+R_{k_2}-\Delta k)a^\prime b^\prime \right]^2 \nonumber \\
 && +\left[(R_{k_1}-R_{k_2}-\Delta k)a^\prime - (R_{k_1}-R_{k_2}+\Delta k)b^\prime \right]^2 .
\end{eqnarray}

The functions $a^\prime$ and $b^\prime$ are given by
\begin{eqnarray}
a^\prime = {\hat \alpha}^\prime \, {e^{2U_{k_1}} + e^{2{\tilde U}_{\nu_1}} \over e^{{\tilde U}_{\nu_1}} }
{e^{2U_{k_1}} + e^{2{\tilde U}_{\nu_2}} \over e^{{\tilde U}_{\nu_2}} }
{e^{{\tilde U}_{\nu_3}} \over e^{2U_{k_1}} + e^{2{\tilde U}_{\nu_3}}}
{e^{2U_{k_1}} + e^{2{\tilde U}_{\nu_4}} \over e^{{\tilde U}_{\nu_4}}}
{e^{{\tilde U}_{\nu_5}} \over e^{2U_{k_1}} + e^{2{\tilde U}_{\nu_5}}},
\end{eqnarray}
\begin{eqnarray}
b^\prime = {\hat \beta}^{\prime} \, {e^{{\tilde U}_{\nu_1}} \over e^{2U_{k_2}} + e^{2{\tilde U}_{\nu_1}} }
{e^{{\tilde U}_{\nu_2}} \over e^{2U_{k_2}} + e^{2{\tilde U}_{\nu_2}}}
{e^{2U_{k_2}} + e^{2{\tilde U}_{\nu_3}} \over e^{{\tilde U}_{\nu_3}} }
{e^{{\tilde U}_{\nu_4}} \over e^{2U_{k_2}} + e^{2{\tilde U}_{\nu_4}} }
{e^{2U_{k_2}} + e^{2{\tilde U}_{\nu_5}} \over e^{{\tilde U}_{\nu_5}} }.
\end{eqnarray}

In this way we have constructed an exact solution to the 5D Einstein-Maxwell equations. It is useful
to present the metric of the found solution in more concise and tractable form. For this purpose we shall introduce
the following function $\Omega$ defined by
\begin{eqnarray}
&&\rho^{-1}\partial_{\rho}\Omega = {3\over 4}\left[(\partial_{\rho}u_{0})^2 - (\partial_{z}u_{0})^2\right],\\
&&\rho^{-1}\partial_{z}\Omega = {3\over 2}\left[\partial_{\rho}u_{0}\partial_{z}u_{0} \right]. \nonumber
\end{eqnarray}

The 5D metric then can be written in  the form
\begin{eqnarray}
ds^2 = {g^E_{00}\over W}\left(dt + \omega d\psi \right)^2 + {{\tilde g}^E_{\psi\psi}\over W}d\psi^2 +
{Y^3\over W} g^E_{\rho\rho}\left(d\rho^2 + dz^2\right) + W^2 g^{E}_{\phi\phi}d\phi^2
\end{eqnarray}
where the 5D metric
\begin{eqnarray}\label{vacuumsolution}
ds^2_E = g^E_{00}\left(dt + \omega d\psi \right)^2 + {\tilde g}^E_{\psi\psi} d\psi^2 +
g^E_{\rho\rho}\left(d\rho^2 + dz^2\right) +  g^{E}_{\phi\phi}d\phi^2
\end{eqnarray}
is a solution to the 5D vacuum Einstein equations generated
from the seed (\ref{seed}) and the functions $g^E_{\rho\rho}$ and $Y$ are given by
\begin{eqnarray}
&&g^{E}_{\rho\rho} = e^{2({\hat \Gamma}_2 +\Omega)}= C_{2} {Ae^{2(H_2+\Omega)}\over (R_{l_1} + R_{l_2} )^2 - (\Delta l)^2 },  \\ \nonumber \\
&&Y = e^{2({\hat \Gamma}_{1} - {1\over 3}\Omega)}= Y_{0} {W_{1}\over (R_{k_1} + R_{k_2})^2 - (\Delta k)^2}\,  e^{2(H_1-{1\over 3}\Omega)} ,
\end{eqnarray}
where $C_2$ and $Y_{0}$ are constants. The constant $Y_{0}$ can be chosen such as the vacuum Einstein solution (\ref{vacuumsolution}) to be recovered for $k_1=k_2$, namely
\begin{eqnarray}\label{YCON}
Y_{0}= {1\over  (1+ {\hat \alpha}^\prime {\hat \beta}^\prime)^2 } .
\end{eqnarray}
It is worth mentioning that this choice is consistent with the asymptotic flatness as we will see in the next section.

Let us give the explicit form of the function $H_{1}-{1\over 3}\Omega$ and $H_2 + \Omega$  which can be easily found from the formulas presented above
\begin{eqnarray}
H_1 - {1\over 3}\Omega &=& \gamma_{k_1,k_1} - 2\gamma_{k_1,k_2} + \gamma_{k_2,k_2}  + \left(\gamma_{k_1,\nu_1} -\gamma_{k_2,\nu_1} \right) + \left(\gamma_{k_1,\nu_2} -\gamma_{k_2,\nu_2} \right)  \nonumber \\ &&   - \left(\gamma_{k_1,\nu_3} -\gamma_{k_2,\nu_3} \right) +
\left(\gamma_{k_1,\nu_4} -\gamma_{k_2,\nu_4} \right) - \left(\gamma_{k_1,\nu_5} -\gamma_{k_2,\nu_5} \right) ,
\end{eqnarray}

\begin{eqnarray}
H_2 + \Omega &=& \gamma_{l_1,l_1} - 2\gamma_{l_1,l_2} + \gamma_{l_2,l_2}  + \left(\gamma_{l_1,\nu_1} -\gamma_{l_2,\nu_1} \right) - \left(\gamma_{l_1,\nu_2} -\gamma_{l_2,\nu_2} \right)  \nonumber \\ &&   + \left(\gamma_{l_1,\nu_3} -\gamma_{l_2,\nu_3} \right) +
\left(\gamma_{l_1,\nu_4} -\gamma_{l_2,\nu_4} \right) - \left(\gamma_{l_1,\nu_5} -\gamma_{l_2,\nu_5} \right)
\nonumber \\ && + \gamma_{\nu_1,\nu_1} + \gamma_{\nu_2,\nu_2} + \gamma_{\nu_3,\nu_3} + \gamma_{\nu_4,\nu_4} + \gamma_{\nu_5,\nu_5} \nonumber \\ && +  \gamma_{\nu_1,\nu_2} - \gamma_{\nu_1,\nu_3} + 2\gamma_{\nu_1,\nu_4}
  - 2\gamma_{\nu_1,\nu_5}   - 2\gamma_{\nu_2,\nu_3} + \gamma_{\nu_2,\nu_4} - \gamma_{\nu_2,\nu_5} \nonumber \\ &&
- \gamma_{\nu_3,\nu_4}  + \gamma_{\nu_3,\nu_5} - 2\gamma_{\nu_4,\nu_5} .
\end{eqnarray}

Clearly, in the limit $\Delta k\to 0$ our Einstein-Maxwell solution reduces to the vacuum solution (\ref{vacuumsolution}). Let us  note that the above Einstein-Maxwell solution can be also generated as a non-linear superposition of two solutions (\ref{vacuumsolution})  with appropriate parameters by
using the solution generating method presented in \cite{Yazadjiev3} and this could be easily checked.

\section{Analysis  of the solution}

Since the solution is invariant under the shifts in $z$-direction we can choose, without loss of generality,
$l_1=\sigma$ and $l_2=-\sigma$ ($\Delta l=2\sigma$). This can be achieved just by performing the shift $z\to z+z_0$, where $z_0={1\over 2}(l_1 + l_2)$. The translated parameters $\nu_{i}$,$k_i$ will be denoted as follows
\begin{eqnarray}\label{newparameterization}
\nu_1 \to \lambda \sigma, \,\, \nu_2 \to \delta_1 \sigma ,\,\, \nu_3 \to \delta_2 \sigma, \,\, \nu_4 \to \eta_1 \sigma ,
\,\, \nu_5 \to \eta_2 \sigma, \,\, k_1\to \sigma\zeta_1, \,\, k_2\to \sigma\zeta_2 .
\end{eqnarray}

We also define the new parameters $\alpha$, $\beta$ and $\alpha^\prime$, $\beta^\prime$ by
\begin{eqnarray}\label{newparameterization1}
{\hat \alpha}= {\alpha\over 2\sqrt{\sigma} }, \,\,\, {\hat \beta}= 2\sqrt{\sigma}\beta ,\\ \nonumber \\
{\hat \alpha}^\prime= {\alpha^\prime\over \sqrt{\sigma} }, \,\,\, {\hat \beta}^\prime= \sqrt{\sigma}\beta^\prime \nonumber.
\end{eqnarray}

In this parametrization the vacuum Einstein solution (\ref{vacuumsolution}) coincides with the Iguchi-Mishima solution \cite{IgichiMishima2}. From now on we will work with the parametrization given by (\ref{newparameterization})
and (\ref{newparameterization1}).

The overall investigation of the solution shows that we must chose the following ordering of the parameters
\begin{eqnarray} \label{BOUNDS}
\zeta_2 \le \zeta_1 <-1<\eta_1 <\eta_2<1 <\delta_1 <\delta_2 < \lambda
\end{eqnarray}
in order for the solution to describe concentric rotating dipole black rings.

The careful investigation of the functions  $W$ and $Y$ shows that they have potential singularities at $z=k_1$ and $z=k_2$. In order to eliminate these singularities from $W$ and $Y$ we must impose the following constraints
\begin{eqnarray}
\alpha^\prime = \pm \sqrt{ {(\delta_2 -\zeta_1)(\eta_2-\zeta_1) \over 2 (\lambda- \zeta_1)(\delta_1-\zeta_1)(\eta_1-\zeta_1)}  } ,
\end{eqnarray}
\begin{eqnarray}
\beta^\prime = \pm \sqrt{ 2 (\lambda- \zeta_2)(\delta_1-\zeta_2)(\eta_1-\zeta_2)  \over (\delta_2 -\zeta_2)(\eta_2-\zeta_2)}.
\end{eqnarray}

In order to eliminate the potential singularities in $g^{E}_{00}$ at $z=\sigma$ and $z=-\sigma$
we must impose the constraints
\begin{eqnarray}
\alpha= \pm \sqrt{{2(\delta_1-1)(1-\eta_2)\over (\lambda-1)(\delta_2-1)(1-\eta_1)}}  ,\\  \nonumber \\
\beta= \pm \sqrt{{(1+\lambda)(1+ \eta_1)(1+\delta_2)\over 2(1+\delta_1)(1+\eta_2)}},
\end{eqnarray}
which coincide with those imposed in \cite{IgichiMishima2}.

\subsection{Asymptotic behaviour }

In order to study the asymptotic behaviour of the solution
we introduce the asymptotic coordinates $r$ and $\theta$ defined by
\begin{eqnarray}
&&\rho= \sigma \left[\left({r^2\over 2\sigma} + \lambda+\eta_1-\eta_2 + \delta_2-\delta_1 \right)^2-1\right]^{1/2}\sin2\theta ,\\
&&z= \sigma \left({r^2\over 2\sigma} + \lambda+\eta_1-\eta_2 + \delta_2-\delta_1 \right) \cos2\theta .\nonumber
\end{eqnarray}

In this coordinates the asymptotic behaviour of metric functions is the following
\begin{eqnarray}\label{ASYMP}
&&g^{E}_{00} \approx -1 + {2\sigma\over r^2} \left[(\delta_2-\delta_1) +2 {1 + \alpha^2-\alpha^2\beta^2\over (1+ \alpha\beta)^2}\right],\nonumber \\ \nonumber \\
&& \omega \approx -{2\sqrt{\sigma}\alpha\over 1+ \alpha\beta} + C_{\omega} ,\nonumber
\\\nonumber \\
&&g^{E}_{t\psi}\approx - {\sin^2\theta \over r^2}{4\sigma^{3/2}\over (1+ \alpha\beta)^3 } \left[2\alpha^3 + 2\alpha(1-\alpha^2\beta^2) +
\alpha(1+\lambda + \delta_2 -\delta_1 + \eta_1-\eta_2)(1 + \alpha\beta)^2  \nonumber \right. \\ && \left. -2\beta(1+\alpha\beta)^2 -2\alpha^2\beta(1+\alpha\beta) \right] , \nonumber \\ \nonumber \\
&&g^{E}_{\rho\rho}\approx {\sqrt{2}(1+ \alpha\beta)^2C_{2}\over r^2} ,\\ \nonumber \\
&&g^{E}_{\psi\psi}\approx r^2 \sin^2\theta ,\nonumber \\ \nonumber \\
&&g^{E}_{\phi\phi}\approx r^2 \cos^2\theta ,\nonumber \\ \nonumber \\
&& W\approx 1 - {2\sigma\Delta \zeta\over r^2 (1+ \alpha^\prime\beta^\prime)^2 }\left[1 -{\alpha^\prime}^2{\beta^\prime}^2  +
2\Delta \zeta {\alpha^\prime}^2\right], \nonumber \\ \nonumber \\
&& Y\approx Y_{0} (1+ \alpha^\prime \beta^\prime )^2 \nonumber .
\end{eqnarray}

For the gauge potential we find
\begin{eqnarray}
&&A_{\phi} \approx  A^0_\phi + \sqrt{\sigma} {2\sqrt{3}\Delta \zeta\alpha^\prime\over 1+ \alpha^\prime\beta^\prime } +
\sigma {\cos^2\theta \over r^2} {2\sqrt{3} \Delta \zeta \over 1+ \alpha^\prime\beta^\prime }
\left[\beta^\prime   + 2\alpha^\prime (2\zeta_2-\zeta_1) \right.\nonumber \\ && \left.+ 2 \alpha^\prime (-\lambda -\delta_1 + \delta_2 -\eta_1 + \eta_2)
+ {6\Delta \zeta {\alpha^\prime}^2\beta^\prime\over 1 + \alpha^\prime\beta^\prime } - {4(\Delta \zeta)^2{\alpha^\prime}^3\over (1+ \alpha^\prime\beta^\prime)^2  } \right].
\end{eqnarray}
Here we choose
\begin{eqnarray}
A^0_\phi =- \sqrt{\sigma} {2\sqrt{3}\Delta \zeta\alpha^\prime\over 1+ \alpha^\prime\beta^\prime }
\end{eqnarray}
and the gauge potential has the standard dipole asymptotic behaviour $A_{\phi} \sim \cos^2\theta/r^2$.

In order that our solution  be asymptotically flat we must impose the following conditions
\begin{eqnarray}
&&C_{\omega}= {2\sqrt{\sigma}\alpha\over 1+ \alpha\beta} ,\\\nonumber \\
&& C_{2} ={1\over \sqrt{2}(1+ \alpha\beta)^2} ,\\ \nonumber \\
&& Y_{0}= {1\over (1+ \alpha^\prime \beta^\prime )^2} .
\end{eqnarray}

The first two conditions coincide with the conditions derived in \cite{IgichiMishima2} while the third condition is consistent
with the condition (\ref{YCON}).
In what follows we shall show that  the periodicities of the angles $\phi$ and $\psi$ can be chosen to be the canonical ones $\Delta\phi=\Delta\psi=2\pi$ and as a consequence we find that the spacetime described by our solution is indeed asymptotically flat.

\subsection{Rod structure and balance conditions}

\begin{figure}
\begin{center}
      \includegraphics[width=12cm]{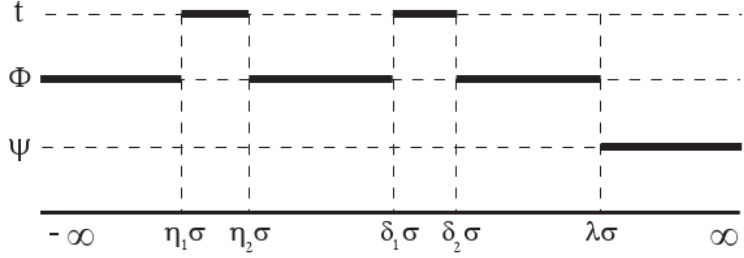}
\caption{Rod structure of the dipole di-ring} \label{Rodstructure}
        \end{center}
\end{figure}

The rod structure is shown in the figure and is described as follows.

* Semi-infinite rod $[\lambda\sigma,+\infty)$ which is a source of the $\psi\psi$ part of the metric.
In order to eliminate the possible conical singularities at the location of the rod the coordinate $\psi$
must be periodic with a period

\begin{eqnarray}
{\Delta \psi}_{Rod[\lambda\sigma,+\infty)}= 2\pi\lim_{\rho\to 0}\sqrt{\rho^2 g_{\rho\rho}\over g_{\psi\psi}}= 2\pi .
\end{eqnarray}

*One semi-infinite rod $(-\infty,\eta_1\sigma]$ and two finite rods $[\eta_{2}\sigma,\delta_1\sigma]$, $[\delta_2\sigma,\lambda\sigma]$ which are sources of the $\phi\phi$ part of the metric.

Let us first consider the semi-infinite rod $(-\infty,\eta_1\sigma]$. Then the regularity condition
gives the period
\begin{eqnarray}
{\Delta \phi}_{Rod(-\infty,\eta_1\sigma]}= 2\pi\lim_{\rho\to 0}\sqrt{\rho^2 g_{\rho\rho}\over g_{\phi\phi}}=2\pi .
\end{eqnarray}

For the first finite rod the regularity condition fixes the period
\begin{eqnarray}
{\Delta\phi}_{Rod[\eta_2\sigma,\delta_1\sigma]}&=&\left({Y\over W}\right)_{Rod[\eta_2\sigma,\delta_1\sigma]}^{3/2} \left({2\pi\over 1 +\alpha\beta}\right)
\sqrt{(\lambda+1)(\delta_1-1)(\delta_2+1)(\delta_1-\eta_2)(\delta_2-\eta_1)\over (\lambda-1)(\delta_1+1)(\delta_2-1)(\delta_1-\eta_1)(\delta_2-\eta_2) } \nonumber \\
&&\times \left(\lambda-\eta_2\over \lambda-\eta_1\right) \left[1 +  \alpha\beta{(\lambda-1)(\delta_1+1)(\delta_2-1)\over (\lambda+1)(\delta_1-1)(\delta_2+1)} \right],
\end{eqnarray}
where
\begin{eqnarray}
\left({Y\over W}\right)_{Rod[\eta_2\sigma,\delta_1\sigma]}= \left[ { 1 +
\alpha^\prime\beta^\prime {(\lambda-\zeta_1)(\delta_1-\zeta_1)  \over (\delta_2-\zeta_1) } { (\delta_2 -\zeta_2)\over (\lambda -\zeta_2)(\delta_1-\zeta_2)}   \over  1+ \alpha^\prime\beta^\prime } \right]^2
\left[{(\lambda-\zeta_2)(\delta_1-\zeta_2) (\delta_2-\zeta_1) \over (\lambda-\zeta_1)(\delta_1-\zeta_1)(\delta_2-\zeta_2)} \right] .
\end{eqnarray}

The regularity condition for the second finite rod gives
\begin{eqnarray}
{\Delta \phi}_{Rod[\delta_2\sigma,\lambda\sigma]} = \left({Y\over W}\right)_{Rod[\delta_2\sigma,\lambda\sigma]}^{3/2} \left(2\pi\over 1 + \alpha\beta\right)
\sqrt{(\lambda+1)(\lambda-\delta_2)\over (\lambda-1)(\lambda-\delta_1) }  \nonumber \\
\times    \left({\lambda-\eta_2\over \lambda-\eta_1}\right)\left[1 + \alpha\beta{\lambda-1\over \lambda+1 }\right] ,
\end{eqnarray}
where
\begin{eqnarray}
\left({Y\over W}\right)_{Rod[\delta_2\sigma,\lambda\sigma]} = \left[ {1 + \alpha^\prime\beta^\prime {\lambda-\zeta_1\over \lambda-\zeta_2}  \over 1 + \alpha^\prime\beta^\prime} \right]^2 \left({\lambda -\zeta_2\over \lambda-\zeta_1}\right) .
\end{eqnarray}

The balancing conditions are
\begin{eqnarray}
{\Delta\phi}_{Rod[\eta_2\sigma,\delta_1\sigma]}=2\pi ,\\\nonumber \\
{\Delta \phi}_{Rod[\delta_2\sigma,\lambda\sigma]}=2\pi .
\end{eqnarray}

The parameters can be adjusted so that  the balancing conditions  be satisfied.

\subsection{Horizons }

There are two horizons located at $\rho=0$ for $z\in [\eta_1\sigma,\eta_2\sigma]$ and $z\in[\delta_1\sigma,\delta_2\sigma]$. Both horizons have $S^2\times S^1$ topology. The
first horizon corresponds to the outer black ring and the second to the inner black ring.

\subsubsection{Outer black ring horizon}

The area of the horizon is

\begin{eqnarray}
{\cal A}_{1}= {4\sqrt{2}\pi^2 \sigma^{3/2}Y^{3/2}_{h}\over (1+ \alpha\beta)^2 }
\left[2(1+ \alpha\beta) +\alpha^2(1-\eta_1) -2\alpha\beta {(\delta_1 +1)(\eta_2 +1)\over (\lambda+1)(\delta_2 +1)} \right] \nonumber \\ \times \sqrt{(\lambda +1) (\delta_2+1)(\delta_2-\eta_1)   \over (\delta_1 +1)(\eta_2+1)(\delta_1-\eta_1)(1-\eta_1)} \, {(\eta_2-\eta_1)^2\over (\lambda-\eta_1)}
\end{eqnarray}
where
\begin{eqnarray}
Y_{h}= \left[ { 1+ \alpha^{\prime}\beta^\prime   {(\lambda -\zeta_1)(\delta_1-\zeta_1)(\delta_2-\zeta_2)(\eta_2-\zeta_2)\over (\delta_2 -\zeta_1)(\eta_2-\zeta_1)(\lambda-\zeta_2)(\delta_1-\zeta_2) }    \over     1 + \alpha^\prime\beta^\prime}  \right]^2
{ (\lambda-\zeta_2) (\delta_1-\zeta_2)(\delta_2 -\zeta_1)(\eta_2-\zeta_1)  \over  (\lambda -\zeta_1)(\delta_1-\zeta_1)(\delta_2-\zeta_2)(\eta_2-\zeta_2)  }.
\end{eqnarray}

The temperature of the horizon can be calculated via the surface gravity. There is alternative way to find the temperature by using the formula derived in \cite{HolandsYazadjiev1}. This formula relates the length $l_h$ of the horizon rod with the surface gravity $\kappa$ and area of the horizon, namely

\begin{eqnarray}
l_{h}= {1\over 4\pi^2} \kappa {\cal A}.
\end{eqnarray}
In our case, this formula gives
\begin{eqnarray}
T_1 = {2\pi} {\sigma (\eta_2-\eta_1)\over {\cal A}_{1} }.
\end{eqnarray}

The angular velocity of the horizon is given by
\begin{eqnarray}
\Omega^{h}_1 =  {1 + \alpha\beta\over 2\sqrt{\sigma} } {  2\beta(\delta_1 + 1)(1+\eta_2) - \alpha (1-\eta_1)(\lambda+1)(\delta_2 +1) \over 2\alpha\beta (\delta_1+1)(1+\eta_2) -(\lambda+1)(\delta_2+1)\left[\alpha^2(1-\eta_1) + 2\alpha\beta +2 \right]}.
\end{eqnarray}

\subsubsection{Inner black ring horizon}

The area of the horizon is
\begin{eqnarray}
{\cal A}_{2} = {4\sqrt{2}\pi^2 \sigma^{3/2}Y^{3/2}_{h}\over (1+ \alpha\beta)^2 }
\sqrt{\left[\alpha\beta(\lambda-1) + (\lambda+1)\right] \left[\alpha\beta(\delta_2-1) + (\delta_2+1)\right]} \nonumber \\ \times \sqrt{{(\delta_1-\eta_1)\over (\lambda^2-1)(\delta^2_2-1)(\lambda-\delta_1)(\delta_2-\eta_2) }}
{(\lambda-\eta_2)(\delta_2-\delta_1)^2\over (\lambda- \eta_1) }
\end{eqnarray}
where
\begin{eqnarray}
Y_{h} =  \left[{1 + \alpha^\prime \beta^\prime {(\lambda -\zeta_1)(\delta_2-\zeta_2)\over (\lambda-\zeta_2)(\delta_2-\zeta_1) }   \over  1 +  \alpha^\prime \beta^\prime} \right]^2 { (\lambda-\zeta_2)(\delta_2 -\zeta_1)  \over (\lambda -\zeta_1)(\delta_2-\zeta_2)} .
\end{eqnarray}

The temperature of the horizon is
\begin{eqnarray}
T_2= 2\pi {\sigma (\delta_2-\delta_1)\over {\cal A}_2 }.
\end{eqnarray}

For the angular velocity we find
\begin{eqnarray}
\Omega^{h}_2= - {1\over \sqrt{\sigma}} { 2\beta(1+ \alpha\beta)  \over \left[\alpha\beta (\lambda-1) +(\lambda+1) \right]
\left[\alpha\beta(\delta_2-1) + (\delta_2+1) \right] }.
\end{eqnarray}

\subsection{Masses, angular momenta and dipole charges}

\subsubsection{ADM mass and angular momentum}

The ADM mass\footnote{The expression for the ADM mass of the vacuum black di-ring given in \cite{IgichiMishima2} seems to be incorrect. } and angular momentum can be found from the asymptotic behaviour of the metric (\ref{ASYMP}) which gives
\begin{eqnarray}
M_{ADM} &=& {3\pi \over 4}\sigma {\Delta \zeta\over (1 + \alpha^\prime\beta^\prime)^2}\left[{\alpha^\prime}^2{\beta^\prime}^2 -1 -
2\Delta \zeta {\alpha^\prime}^2 \right]
 \\ &&+ {3\pi\over 4 }\sigma \left[(\delta_2-\delta_1) + 2 {1+ \alpha^2 -\alpha^2\beta^2\over (1+\alpha\beta)^2} \right],\nonumber  \\\nonumber
\\
J_{\psi} &=& {\pi \sigma^{3/2} \over (1+ \alpha\beta)^3 } \left[2\alpha^3 + 2\alpha(1-\alpha^2\beta^2) + \alpha(1+\lambda + \delta_2-\delta_1 + \eta_1 -\eta_2)(1+\alpha\beta)^2 \right. \nonumber \\&& \left. - 2\beta(1+\alpha\beta)^2 - 2 \alpha^2\beta(1+\alpha\beta) \right].
\end{eqnarray}

One can show that the ADM mass is non-negative for parameters satisfying the bounds (\ref{BOUNDS}).

\subsubsection{Komar mass and angular momentum}

As well-known the Komar mass and angular momentum are defined by the integrals
\begin{eqnarray}
M^{K} = -{3\over 32\pi} \int_{\partial\Sigma} \star d{\tilde \xi} ,\,\,\, \,
J^{K} = {1\over 16\pi} \int_{\partial\Sigma}\star d{\tilde K}_{\psi}
\end{eqnarray}
where ${\tilde \xi}$ and ${\tilde K}_{\psi}$ are $1$-forms dual to the timelike Killing vector $\xi$ and the spacelike Killing vector
$K_{\psi}$. Here $\partial\Sigma$ is a boundary of any spacelike hypersurface $\Sigma$. From a physical point of view the Komar integrals
measure the mass and angular momentum contained in $\partial\Sigma$. When $\partial\Sigma$ is a three-sphere at infinity the Komar
integrals coincide with the ADM mass and angular momentum of an asymptotically flat spacetime. When dealing with multi-horizon
configurations the Komar integrals evaluated on the horizon cross sections  are of special interest since they give the
"intrinsic" mass and angular momenta characterizing the black objects in the considered spacetime geometry.  These  quantities for the black rings in our case are the following

\begin{eqnarray}
M^{K}_{1}=&& {3\pi\over 2 }\sigma  {\alpha\beta \over 1+ \alpha\beta }\left( {\eta_2-\eta_1\over \lambda + 1 }\right) \left({\delta_1+1\over \delta_2 +1 } \right) \\
&&\times
\left[{ 1 + \alpha\beta  - \alpha\beta {1 + \eta_2\over \lambda+1}{\delta_1+1\over \delta_2+1} + {1\over 2}\alpha^2(1-\eta_1)   \over  {1\over 4}\alpha^2(1-\eta^2_1)  + \alpha\beta {1-\eta_1\eta_2\over \lambda+1 }{\delta_1 +1\over \delta_2+1}  + \beta^2 {1-\eta^2_2\over (\lambda+1)^2} {(\delta_1+1)^2\over (\delta_2 + 1)^2 }}\right] , \nonumber
\end{eqnarray}

\begin{eqnarray}
J^{K}_1 = {2\over 3\Omega^h_1} \left[ M^{K}_{1} - {3\pi \over 4} \sigma (\eta_2-\eta_1)\right],
\end{eqnarray}

\begin{eqnarray}
M^{K}_2= {3\pi\over 4} \sigma \left({\delta_2-\delta_1\over 1 +\alpha\beta}\right)
{\left(1 + \alpha\beta {\delta_2-1\over \delta_2 +1 }\right)  \left(1 + \alpha\beta {\lambda-1\over \lambda+1} {\delta_1+1\over \delta_1 -1 }\right) \over   \left(1 + \alpha\beta {\lambda-1\over \lambda+1} {\delta_1+1\over \delta_1 -1 } {\delta_2-1\over \delta_2 +1 } \right)} ,
\end{eqnarray}

\begin{eqnarray}
J^{K}_2 = {2\over 3\Omega^h_2} \left[ M^{K}_{2} - {3\pi \over 4} \sigma (\delta_2-\delta_1)\right].
\end{eqnarray}

\subsubsection{Dipole charges}

The dipole charge associated with the 2-sphere of the horizon ${\cal H}_{i}$ is defined by
\begin{eqnarray}
{\cal Q}_{i} ={1\over 4\pi} \oint_{S^2\in {\cal H}_i}F .
\end{eqnarray}

For our solution we find
\begin{eqnarray}
{\cal Q}_{1}= \sqrt{\sigma}{\sqrt{3}\Delta \zeta \alpha^\prime \over 1+\alpha^\prime\beta^\prime}   \left[ { 1 + \alpha^\prime\beta^\prime {\eta_1-\zeta_1\over \eta_2-\zeta_2} \over 1 + \alpha^\prime\beta^\prime {(\eta_1-\zeta_1)(\eta_2-\zeta_2)\over (\eta_2-\zeta_1)(\eta_1-\zeta_2) } }\right] {\eta_2-\eta_1 \over \eta_2-\zeta_1 },
\end{eqnarray}

\begin{eqnarray}
{\cal Q}_{2}= {\sqrt{\sigma }\over 2}{\sqrt{3}\Delta \zeta \beta^{\prime} \over 1 + \alpha^\prime\beta^\prime {(\lambda - \zeta_1)\over (\lambda-\zeta_2)} }  \left[ { 1 + \alpha^\prime\beta^\prime  { (\lambda-\zeta_1)(\delta_2-\zeta_2)\over (\lambda-\zeta_2)(\delta_2-\zeta_1)}  \over 1 + \alpha^\prime\beta^\prime  { (\lambda-\zeta_1)(\delta_1-\zeta_1)(\delta_2-\zeta_2)\over
(\lambda-\zeta_2)(\delta_1-\zeta_2)(\delta_2-\zeta_1)}} \right] {(\delta_2-\delta_1)\over (\lambda -\zeta_2)(\delta_1-\zeta_2)},
\end{eqnarray}
where ${\cal Q}_1$ and ${\cal Q}_2$ are the dipole charges for the outer and the inner black ring, respectively.
The dipole charges for the balance dipole black di-ring can vanish only in the limit when the  length of the corresponding horizon rod becomes  zero. In other words, our solution does not allow a case when only one of the black rings is charged. Physically this is explained by the fact that the dipole magnetic field of each black  ring induces dipole charge on the other.

\subsubsection{Dipole potential }

Let us consider the dual field $H$ defined by
\begin{eqnarray}
H = \star F.
\end{eqnarray}

One can show that there exists a 2-form $B$ such that
\begin{eqnarray}
H=dB.
\end{eqnarray}
Fortunately,  the 2-form $B$ can be found in explicit form and it is given by
\begin{eqnarray}
B_{t\psi}=  e^{-u_{0}} {{\hat \omega}_{k_1,k_2} \over W_1} + C_B
\end{eqnarray}
where $C_{B}$ is a constant. Here we fix the constant by requiring  $B_{t\psi}$ to be zero at spacial infinity which gives
\begin{eqnarray}
C_{B}= {2\sqrt{\sigma}\Delta \zeta\,  \alpha^\prime\over (1+ \alpha^\prime\beta^\prime)}.
\end{eqnarray}

Further we can define the dipole potentials on the black ring horizons as follows
\begin{eqnarray}
\Xi_i = - {\pi\over 2} B_{t\psi}|_{{\cal H}_i}.
\end{eqnarray}
After some algebra we find
\begin{eqnarray}
&&\Xi_1 =\sqrt{\sigma} {\pi\sqrt{3}\over 2 (1 + \alpha^\prime\beta^\prime)}  {\Delta \zeta\over (\eta_1-\zeta_2) }
\left[{\beta^{\prime} + \alpha^\prime {\beta^\prime}^2 - 2\alpha^\prime (\eta_1-\zeta_2) - 2{\alpha^\prime}^2\beta^\prime (\eta_1-\zeta_1)  \over 1 + \alpha^\prime\beta^\prime {\eta_1-\zeta_1\over\eta_1-\zeta_2 } }\right] ,\\ \nonumber \\
&&\Xi_2 = \sqrt{\sigma} {\pi \sqrt{3}\Delta \zeta \alpha^\prime \over 1 + \alpha^\prime\beta^\prime }
\left({\lambda-\delta_2 \over \delta_2-\zeta_1}\right) {1 + \alpha^\prime \beta^\prime {\lambda - \zeta_1\over\lambda-\zeta_2 }  \over 1 + \alpha^\prime\beta^\prime {(\lambda-\zeta_1)(\delta_1-\zeta_2)\over (\delta_2-\zeta_1)(\lambda-\zeta_2)}  }
\end{eqnarray}

\subsubsection{Smarr-like relation}

One can show that the following Smarr-like relations are satisfied:
\begin{eqnarray}
M^{K}_1= {3\over 2}\left(T_1 S_1 + \Omega^{h}_1 J^{K}_1  \right),
\end{eqnarray}

\begin{eqnarray}
M^{K}_2= {3\over 2}\left(T_2 S_2  +  \Omega^{h}_2 J^{K}_2  \right),
\end{eqnarray}

\begin{eqnarray}
M= M^{K}_1 + M^{K}_2 + {1\over 2} \Xi_1 {\cal Q}_1 + {1\over 2} \Xi_2 {\cal Q}_2 ,
\end{eqnarray}
where $S_i={1\over 4} {\cal A}_i$ is the entropy of the $i$-th black ring.

It is natural to assume that the mass of the black di-ring is a weighted homogeneous function of $J^K_1$, $J^K_2$, ${\cal Q}_1$,  ${\cal Q}_2$,  $S_1$ and $S_2$. The corresponding weights can be found from the dimensionality of the quantities. Obviously we have $[S_1]=[S_2]=[M]^{3/2}$, $[J^K_1]=[J^K_2]=[M]^{3/2}$ and  $[{\cal Q}_1]=[{\cal Q}_2]=[M]^{1/2}$.  The Euler's theorem then gives
\begin{eqnarray}
M= {3\over 2}S_1 {\partial M \over\partial S_1 } + {3\over 2}S_2 {\partial M \over\partial S_2 } + {3\over 2}J^K_1 {\partial M \over\partial J^K_1 } + {3\over 2}J^K_2 {\partial M \over\partial J^K_2 } +  {1\over 2}{\cal Q}_1 {\partial M \over\partial {\cal Q}_1 } + {1\over 2}{\cal Q}_2 {\partial M \over\partial {\cal Q}_2 }.
\end{eqnarray}
Comparing this to the Smarr-like relations we find
\begin{eqnarray}
&&{\partial M \over\partial S_1 }=T_1 , \,\,\,\, {\partial M \over\partial S_2 }=T_2 ,\,\,\,\,  \nonumber \\ \nonumber \\&&
 {\partial M \over\partial J^K_1 }=\Omega^h_1, \,\,\,\,{\partial M \over\partial J^K_2 }=\Omega^h_2  , \\ \nonumber \\&&  {\partial M \over\partial {\cal Q}_1 }=\Xi_1 , \,\,\,\,
{\partial M \over\partial {\cal Q}_2 }=\Xi_2 . \nonumber
\end{eqnarray}
As a consequence we obtain the first law
\begin{eqnarray}
\delta M = T_1\delta S_1 + T_2\delta S_2 +
{\Omega}^{h}_1 \delta J^K_1 + {\Omega}^{h}_2 \delta J^K_2 +
\Xi_1 \delta {\cal Q}_1 + \Xi_2 \delta {\cal Q}_2 .
\end{eqnarray}

\subsection{Ergosurfaces and closed timelike curves  }
The preliminary investigation shows that probably there are ergosurfaces for both of the black rings.
Unfortunately the solution is too complicated and explicit analytical description is not possible.
Careful investigation of the behaviour of the metric shows that there are no closed timelike curves and
the solution is free from causal pathologies.

\subsection{Non-uniqueness and uniqueness}
The dipole di-ring solution depends on six
independent parameters. Two of them can be fixed by fixing the conserved asymptotic charges, namely the mass and the angular momentum. Therefore the solution exhibits four-fold continuous non-uniqueness. It is interesting to consider this solution in the spirit of the uniqueness theorems proven in \cite{HolandsYazadjiev1} and \cite{HolandsYazadjiev2}. It is not so difficult to extend the theorems of \cite{HolandsYazadjiev1} and \cite{HolandsYazadjiev2} to the case of multi-horizon solutions. Then, in the spirit of such an extension, we should expect that our solution is uniquely  determined by the rod structure, the mass $M$, angular momentum $J$ and the dipole charges ${\cal Q}_1$ and ${\cal Q}_2$.

\subsection{Limits of the solution}
Let us consider the limits of the balanced dipole di-ring. By removing the outer or the inner  black ring from the configuration we obtain a single dipole black ring. The formal procedure is to take the limit in which
the corresponding horizon rod length goes to zero. As one can see these limits are completely regular.
One can show that there is no regular limit of a merge of the black rings. An interesting question
is whether the dipole black Saturn \cite{Yazadjiev6}(and the vacuum black Saturn \cite{ElvangFigueras} as a particular case) could be obtained from
the dipole di-ring in the limit when the length of the finite rod $[\delta_2\sigma,\lambda\sigma]$ tends to zero.
The careful investigation shows that this limit is singular for the balanced black di-ring and the black Saturn
can not be obtained as a limit of the balanced black di-ring.

\subsection{Physical effects}

The exact solution we have constructed here is too complicated and the complete study of the physics of the dipole di-ring certainly requires numerical calculations. There are however some particular cases which are tractable
and  which can help us to see interesting  physical effects. Below we consider  two particular cases which can be worked out analytically.

The first case we shall consider is  $\eta_2=1$. Let us note that this case is completely regular. For the considered
value of the parameter $\eta_2$ we find

\begin{eqnarray}
&&M^K_{1}= {3\pi\over 2}\sigma , \,\,\, M^K_{2}= {3\pi\over 2}\sigma (\delta_2-\delta_1) , \nonumber \\ \nonumber \\
&&\Omega^h_1=  - {\beta \over \sqrt{\sigma}} {(\delta_1+1)\over (\lambda+1)(\delta_2 +1)} , \,\,\,
\Omega^h_2 = - {1\over \sqrt{\sigma}} {2\beta \over (\lambda+1)(\delta_2 +1) } , \\
&&J^K_1= - {\pi \sigma^{3/2}\over 2\beta} {(1+\eta_1)(\delta_2 +1)(\lambda+1)\over (\delta_1 + 1)}= -2\pi \sigma^{3/2}\beta , \,\,\,
J^K_2= 0 . \nonumber
\end{eqnarray}

In the case under consideration the inner black ring is rotating $\Omega^h_2\ne 0$ despite having zero angular momentum,
$J^K_2=0$. This is a frame-dragging effect which is also observed for the black Saturn \cite{ElvangFigueras}.
The rotating outer black ring drags the spacetime around it and as a consequence the inner black ring rotates too despite having no intrinsic angular momentum.

The second case we will consider is when $\eta_1=-1$. From the general expressions for the physical quantities we obtain

\begin{eqnarray}
&&M^K_1= {3\pi\over 2}\sigma (1+ \alpha^2), \,\,\,  M^K_2= {3\pi \over 4}\sigma (\delta_2-\delta_1), \nonumber \\ \nonumber \\
&&\Omega^h_1=  {1\over 2\sqrt{\sigma}}{\alpha\over 1+ \alpha^2} ,\,\,\, \Omega^h_2=0, \\
&&J^K_1= {\pi\sigma^{3/2}\over \alpha}(1+ \alpha^2)\left[2(1+ \alpha^2)- (1+ \eta_2) \right] ,\,\,\, J^K_2= - \pi\sigma^{3/2}\alpha {\delta_2-\delta_1\over\delta_1-1 }(\lambda-\delta_1).\nonumber
\end{eqnarray}

Here the angular velocity of the inner black ring vanishes, $\Omega^h_2=0$, while its intrinsic angular momentum is non-zero, $J^K_2\ne 0$.
In this case the black rings are counter-rotating such as the intrinsic angular momentum of the inner black
ring cancels the effect of the dragging by the outer black ring.

\section{Conclusion}

In this paper we have discussed the application of the solitonic techniques to a certain  completely  integrable
sector of the 5D Einstein-Maxwell gravity. We have reduced the 5D stationary, axi-symmetric Einstein-Maxwell equations
to two Ernst equations. This reduction allows us to adapt the solitonic techniques developed in four-dimensional gravity
to five dimensions and to construct systematically exact solutions to the considered sector of the 5D Einstein-Maxwell gravity . The application of the solitonic techniques was demonstrated by explicitly constructing the
exact solution which describes  two concentric rotating black rings with dipole charges (the dipole black di-ring). The basic quantities characterizing the solution have been computed and some basic properties of the dipole di-ring have been discussed. The calculated physical quantities open the way to thoroughly studying  the physics of the vacuum and dipole di-ring.

\section*{Acknowledgements}
The author would like to thank the Alexander von Humboldt Foundation for a stipend, and
the Institut f\" ur Theoretische Physik G\" ottingen for its kind hospitality. The partial support by the
Bulgarian National Science Fund under Grant MUF04/05 (MU 408) and VUF-201/06  is also acknowledged.
Part of this work was completed while the author was visiting the Institut f\" ur Theoretische Physik Oldenburg and ZARM Bremen. The author would like to thank J. Kunz and C. Laemmerzahl  for the invitation and kind hospitality.
He would also like to thank D. Doneva for drawing the picture of the dipole di-ring rod structure and to D. Doneva, P. Nedkova and I. Stefanov for reading the manuscript.

\end{document}